\documentclass{eas}
\usepackage{epsf}
\usepackage{graphicx}
\usepackage{rotating}
\def\ga{\mathrel{\mathchoice {\vcenter{\offinterlineskip\halign{\hfil
$\displaystyle##$\hfil\cr>\cr\sim\cr}}}
{\vcenter{\offinterlineskip\halign{\hfil$\textstyle##$\hfil\cr  
>\cr\sim\cr}}}
{\vcenter{\offinterlineskip\halign{\hfil$\scriptstyle##$\hfil\cr
>\cr\sim\cr}}}  
{\vcenter{\offinterlineskip\halign{\hfil$\scriptscriptstyle##$\hfil\cr
>\cr\sim\cr}}}}}
\def\farcs{\hbox{$.\!\!^{\prime\prime}$}}

\begin{document}

\title{Spectral Observations of Envelopes around \\
Stars in Late Stages of Stellar Evolution} 
\runningtitle{Spectral Observations of Stars in Late Stages of Stellar Evolution}
\author{T.\ Bl\"ocker}
\address{Max-Planck-Institut f\"ur Radioastronomie, Bonn, Germany}
\author{K.-H. Hofmann$^{1}$}
\author{G.\ Weigelt$^{1}$}
\begin{abstract}
Interferometric observations of stars in late stages of stellar evolution
and the impact of VLTI observations are discussed. 
Special attention is paid to the 
spectral information that can be derived from these observations
and on the corresponding astrophysical interpretation of the data
by radiative transfer modelling.
It is emphasized that for the robust and non-ambiguous construction of 
dust-shell models it is essential to take diverse and independent observational
constraints into account. 
Apart from matching the spectral energy distribution, the 
use of spatially resolved information plays a crucial role for
obtaining reliable models. 
The combination of long-baseline interferometry data with 
high-resolution single-dish data (short baselines), as obtained, for example, 
by bispectrum speckle interferometry, provide complementary information and 
will improve modelling and interpretation.
\end{abstract}
\maketitle
%
%
\section{Introduction: Late stages of stellar evolution} 
The Very Large Telescope Interferometer (VLTI; see Glindemann, this volume) 
of the European Southern Observatory with its four 8.2\,m unit
telescopes (UTs) and three 1.8\,m auxiliary telescopes (ATs) will certainly
establish a new era of studying the late stages of stellar evolution
within the next few years. 
With a maximum baseline of up to more than 200\,m,
the VLTI will allow observations with 
unprecedented resolution opening up new vistas to a better understanding of
the physics of evolved stars and thus of stellar evolution.

Stars in late stages of stellar evolution 
form therefore an important group among the VLTI
key targets.
During the Red Giant phase, strong  winds 
erode the stellar surfaces leading 
to the formation of circumstellar shells which absorb an increasing fraction
of the visible light and re-emit it in the infrared regime. Accordingly, 
most of these evolved stars are bright infrared objects. The heavy mass loss 
leads to the chemical enrichment of the interstellar medium and therefore 
plays a crucial role
for the understanding of the galactic chemodynamical evolution.

The vast majority of all stars, which
have left their main sequence phase and become Red Giants, 
are of low and intermediate mass and finally evolve along the 
Asymptotic Giant Branch (AGB). 
These luminous, frequently pulsating and
heavily mass-losing AGB stars form an important stellar population which
contributes considerably to light, chemistry and dynamics of galaxies.
The envelopes of AGB stars are the major factories of cosmic dust.
Accordingly, AGB stars are often heavily enshrouded by dust exposing high
fluxes in the infrared and are ideal laboratories to investigate the
interplay between various physical and chemical processes.
Most dust shells around AGB stars are known to be spherically 
symmetric on larger scales, whereas most objects in the immediate successive
stage of proto-planetary nebulae appear 
in axisymmetric geometry. Evidence is growing that 
this break of symmetry takes place already at 
the very end of the AGB evolution. 

Mass loss is also one of the dominant
effects during the evolution of massive stars, virtually leading to an almost
complete peeling of the star. Circumstellar dust shells found around
evolved massive supergiants often show features of non-spherical outflows.
Observing and modelling the circumstellar shells surrounding these stars, 
unveil details of evolution as, for instance, mass-loss rates. 
The presence of fossil shells even gives clues for the evolutionary history.

Dust formation around evolved stars can even continue beyond the Red Giant
stage, as, e.g., in R\,CrB stars or late-type Wolf-Rayet stars.
The production of dust in such hostile environments is still challenging to 
theory. In the instance of Wolf-Rayet stars colliding winds due 
to binarity is one of the favored scenarios. 

High-resolution interferometric observations
reveal details of  disks and dust shells 
of evolved stars and thus improve our knowledge 
of, for example, the mass-loss process and its evolution. In the 
following sections, we discuss high spatial resolution 
observations
and their interpretation by radiative transfer calculations for some 
prominent evolved stars. 

%
\section{Bispectrum speckle interferometry}
\begin{figure}
\begin{minipage}{0.8\textwidth}
\begin{turn}{-90}
\epsfysize=1.00\textwidth
\epsfbox{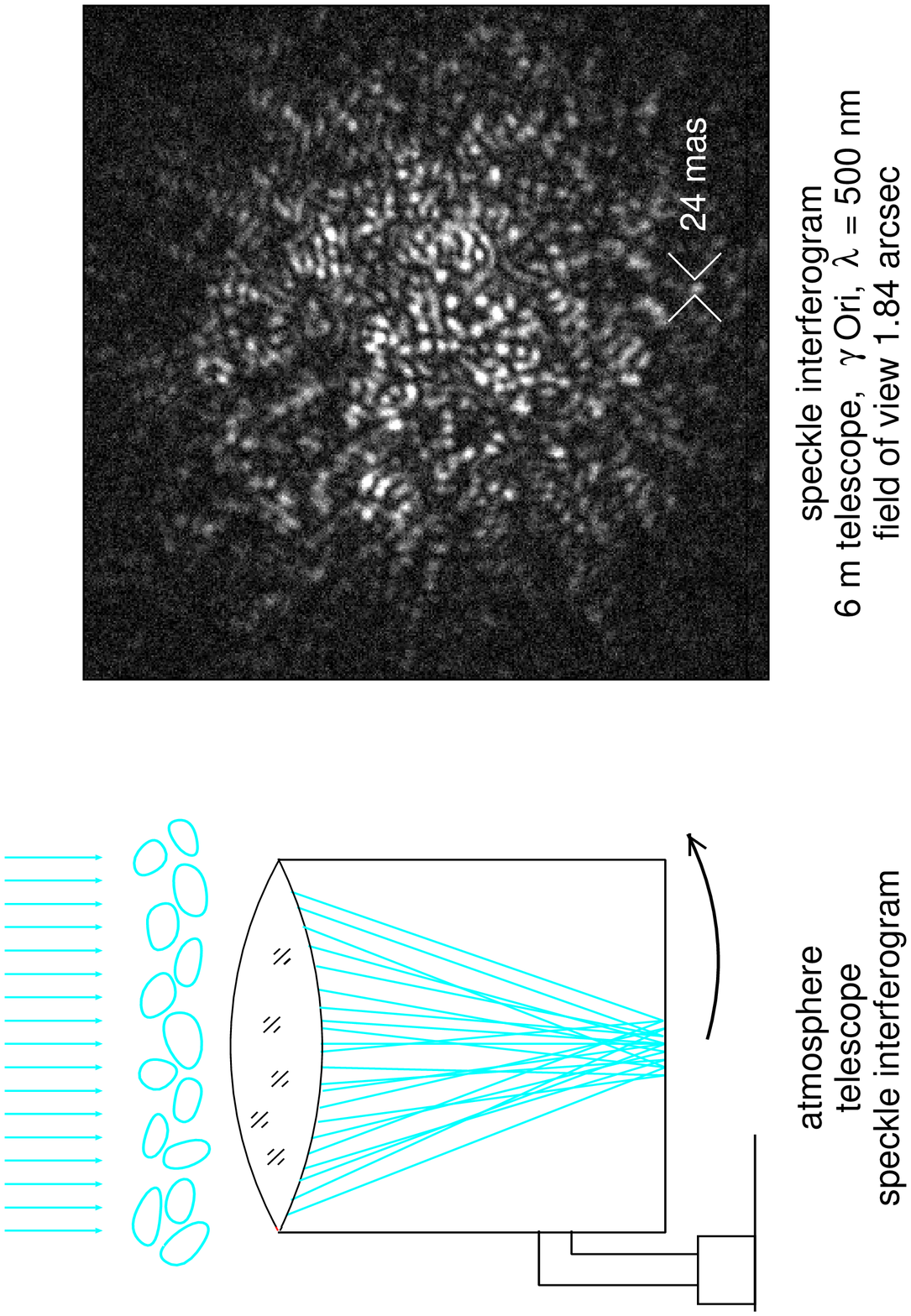}
\end{turn}
\end{minipage}
\hspace*{-5mm}
\begin{minipage}[c]{0.20\textwidth}
\caption{Speckle interferogram of $\gamma$ Ori.}
\label{Fspeckle}
\end{minipage}
\vspace*{-6mm}
\begin{center}
\epsfxsize=40mm
\mbox{\epsffile{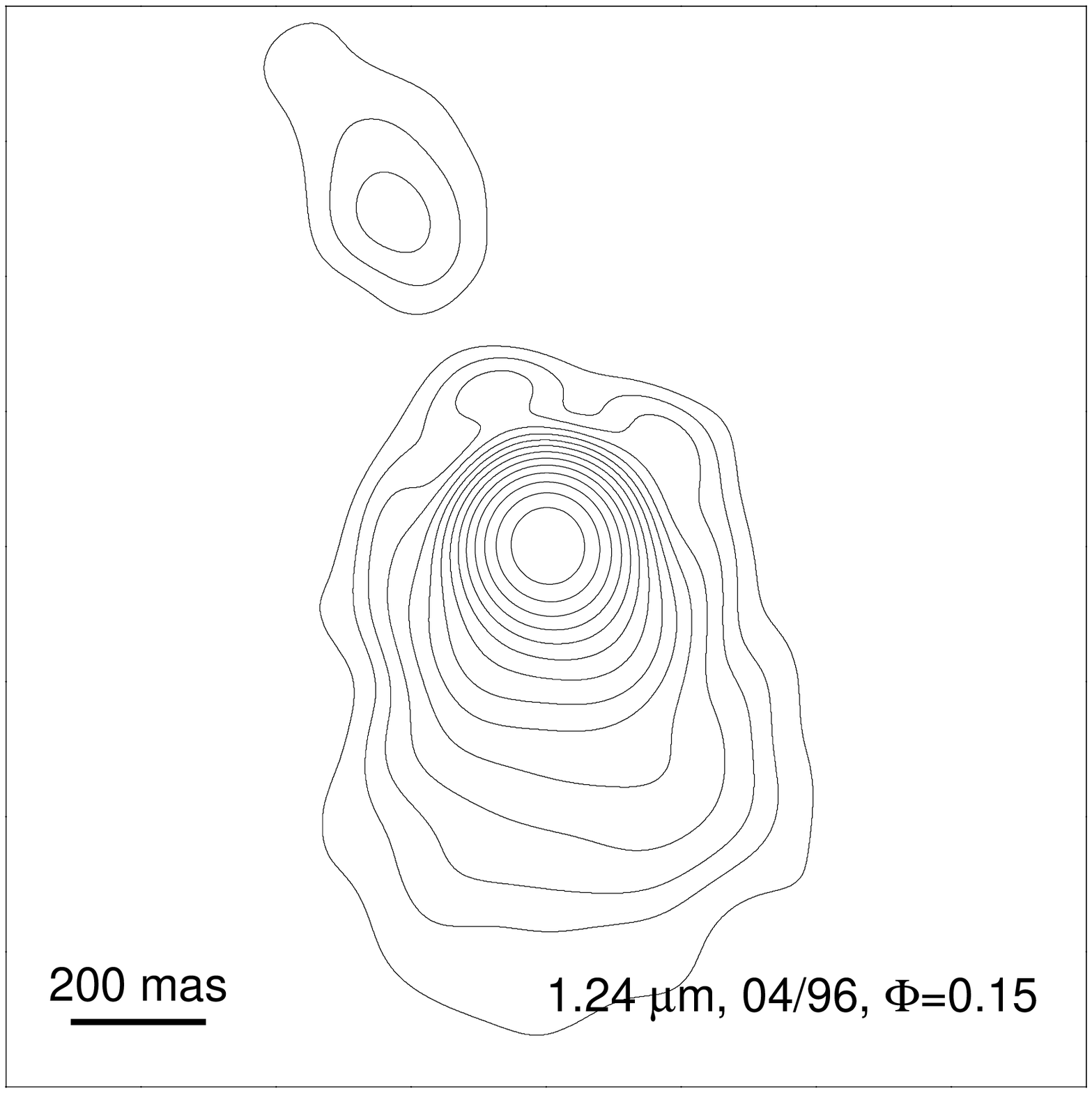}} 
\epsfxsize=40mm
\mbox{\epsffile{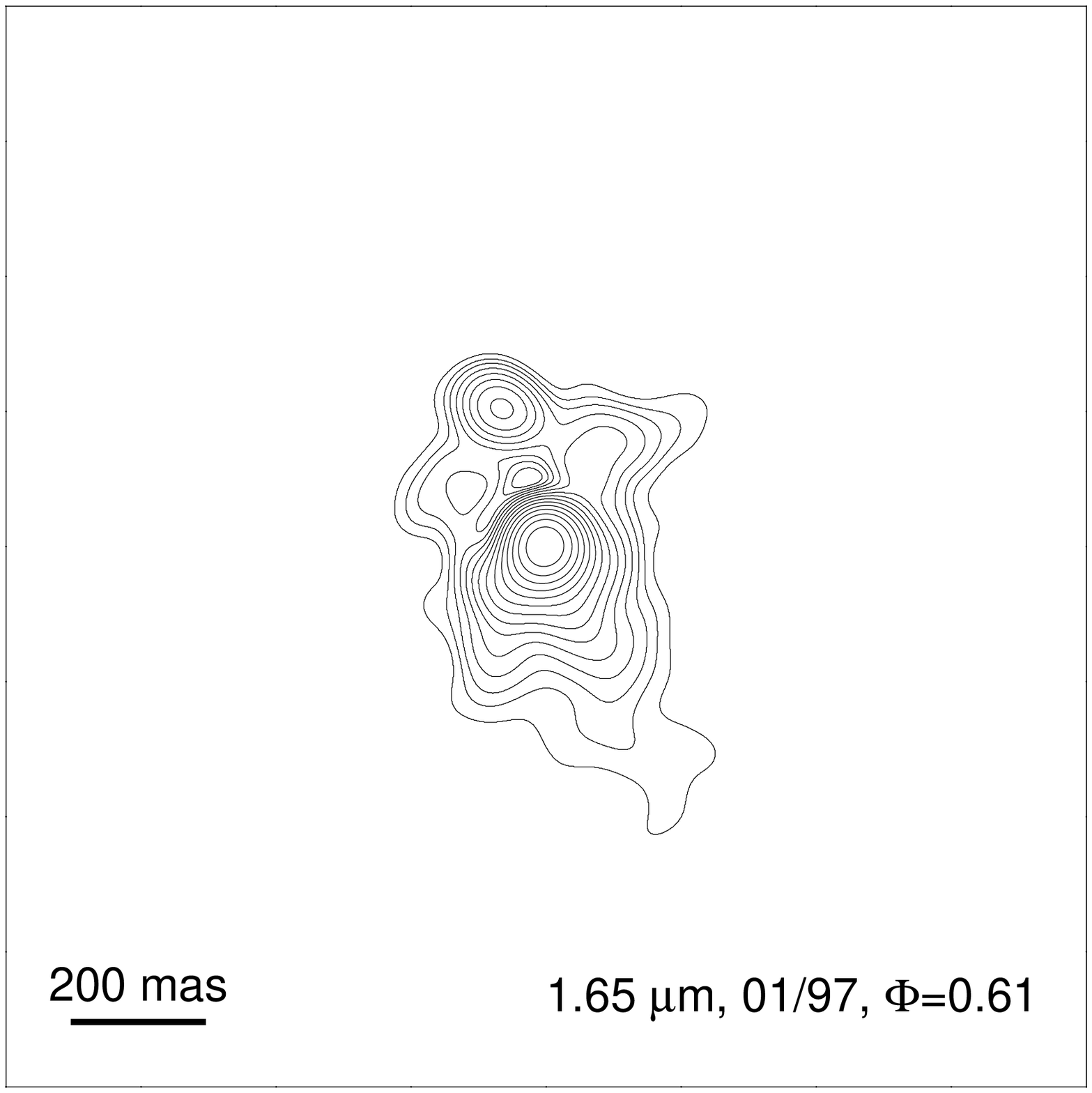}}
\epsfxsize=40mm
\mbox{\epsffile{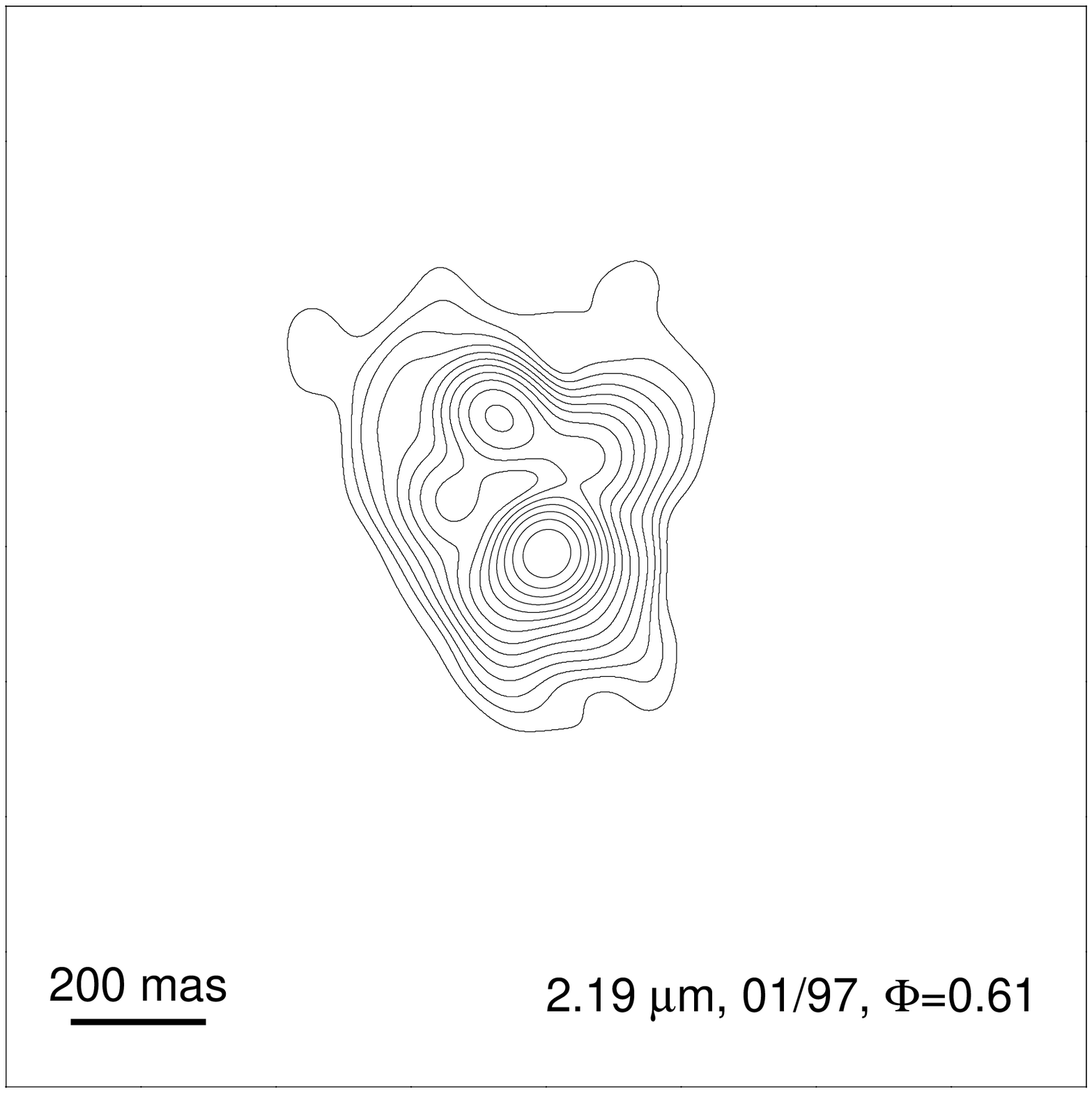}} 
\end{center}
\vspace*{-4mm}
\caption{
Speckle reconstructions of IRC\,+10\,216 in 
$J$ (left, Apr 1996, resolution 146 mas),
$H$ (middle, Jan 1997, resolution 73 mas), and
$K$ (right, Jan 1997, resolution 90 mas)
Contour lines are shown from 0.3\,mag to 4.2\,mag
relative to the peak brightness in steps of 0.3\,mag. 
North is up and east is to the left (Osterbart et al.\ 2000).
}
\label{FJHKcont} 
%
\vspace*{4mm}
\epsfxsize=41.3mm
\mbox{\epsffile{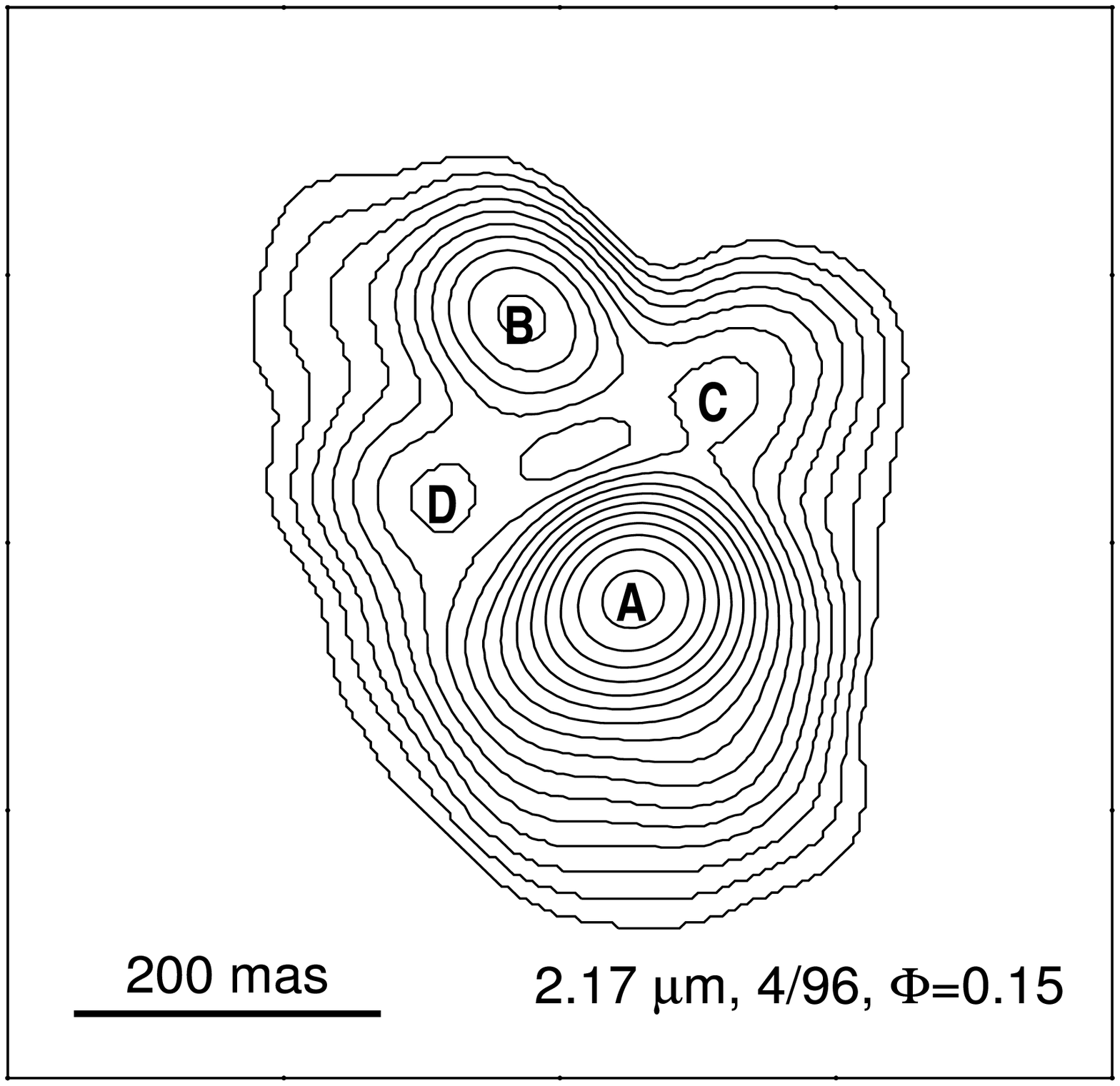}} 
\epsfxsize=40mm
\mbox{\epsffile{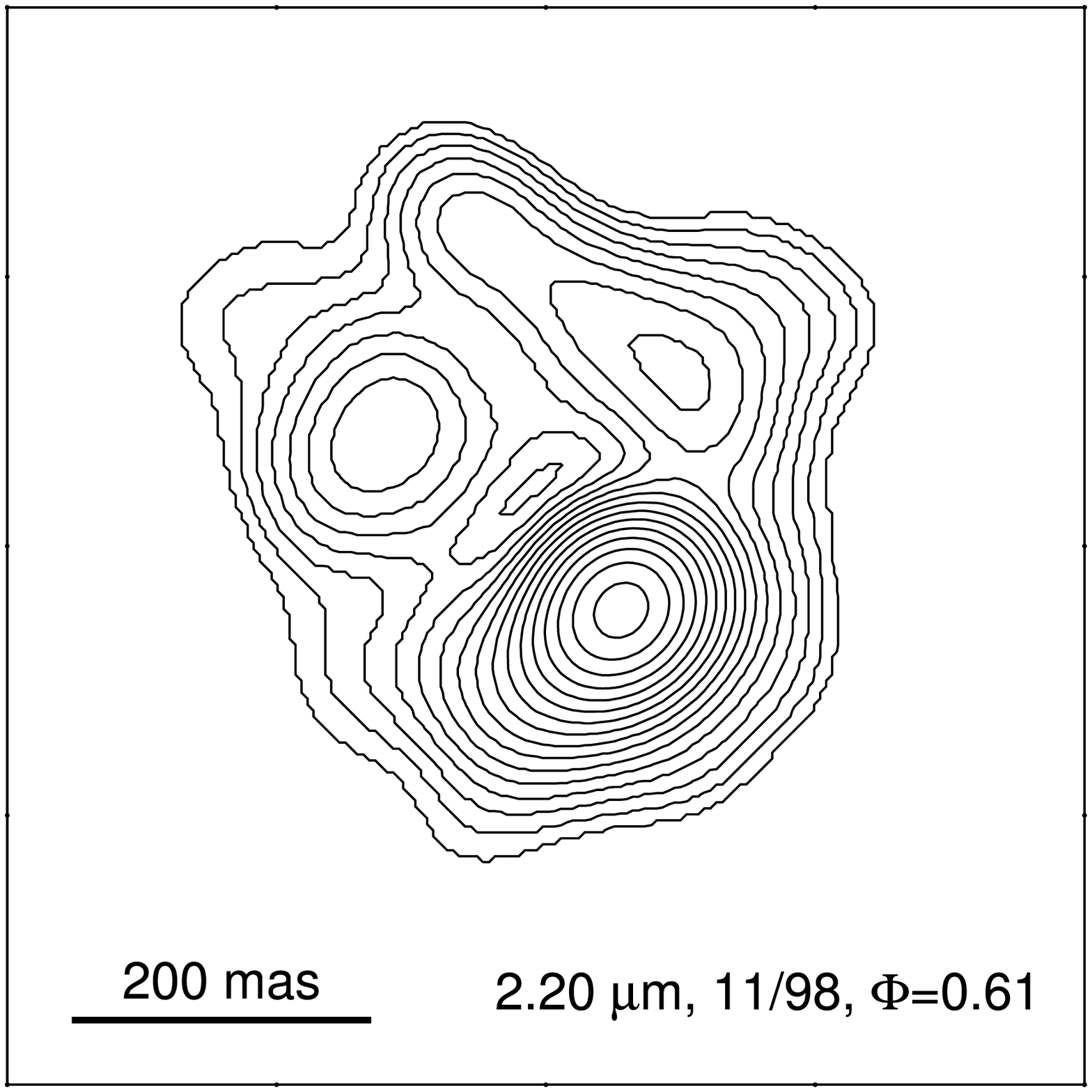}} 
\epsfxsize=40mm
\mbox{\epsffile{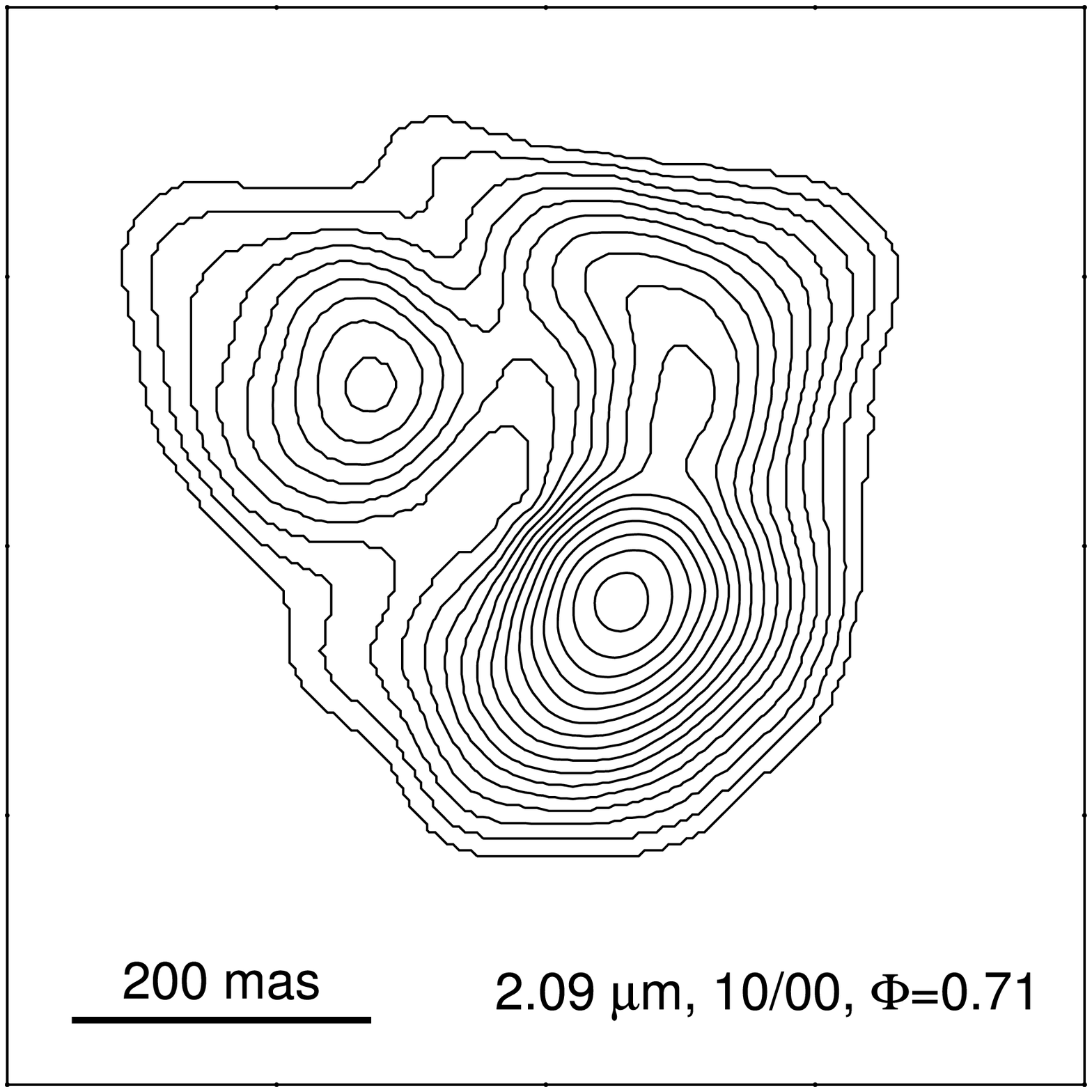}} 
\vspace*{-4mm}
\caption{
$K$-band speckle reconstructions of IRC\,+10\,216
in 1996 (left), 1998 (middle), and 2000 (right).
The resolution is 
82\,mas, 75\,mas, and 73\,mas, resp.
Contour levels are plotted from 0.1\,mag to 3.1\,mag relative
to the peak intensity in steps of 0.2\,mag.
North is up and east is to the left (Weigelt et al.\ 2002).
}
\label{FKima} 
\end{figure}
The refractive index variations in the atmosphere of the earth restrict
the angular resolution of large ground-based telescopes to 
$\sim$ 0\farcs5 which is much worse than the theoretical diffraction limit 
($\sim$0\farcs01 for a 8\,m telescope at optical wavelenghts). However,
the atmospheric image degradation can be overcome and diffraction-limited
images can be obtained either by adaptive 
optics (see, e.g., Tyson et al.\ 2002, Brandner et al.\ 2002) or 
bispectrum speckle interferometry
(Weigelt \cite{Wei77}, Lohmann et al.\ \cite{LohWeiWir83}, Hofmann \& Weigelt 
\cite{HofWei86}).
Speckle interferograms are images recorded with 
exposure times of $\sim 50$\,ms in order to ``freeze'' the atmospheric 
turbulence. They consist of many small bright dots, called speckles,
which are interference maxima of the incident light. The speckles
are typically of the size of the theroretical Airy pattern of the 
aberration-free telescope. 
Fig.~\ref{Fspeckle}. shows a speckle interferogram of $\gamma$ Ori
(6\,m telescope, $\lambda \sim 500$\,nm) for illustration.
Bispectrum speckle interferometry
consists, in principle, of four steps: (i)  calculation of 
the average bispectrum of all speckle interferograms;
(ii) compensation of the photon bias in the bispectrum; 
(iii) compensation of the speckle transfer function which can be 
derived from the speckle interferograms of a point source; and 
(iv) derivation of modulus and phase of the object Fourier transform 
from the bispectrum.  
In other words, bispectrum speckle interferometry with a 
single large telescope 
covers simultaneously all baselines up to the single telescopes's aperture 
size. 
%
%
%
%
\section{Interpreting the observations: Radiative transfer models}
%
%
As discussed above, evolved stars are often surrounded by dust shells.
An appropriate tool to interpret  interferometric observations of dust shells 
are radiative transfer calculations. In order to handle the numerical 
effort to solve this complex problem, various assumptions have usually to be
made with regard to the dust-shell geometry (spherical/2d/3d), 
dust formation (instantaneous/chemical network), and hydrodynamics 
(stationary/time-dependent). For example, 
the spherical radiative transfer problem can be solved by 
utilizing the self-similarity and scaling behaviour of IR emission from 
radiatively heated dust (Ivezi\'c \&  Elitzur \cite{IveEli97}).
To tackle this problem 
including absorption, emission and scattering, several properties of the 
central source and its surrounding envelope are required,
viz. (i) the spectral shape of the central source's
radiation; (ii) the dust properties,
i.e.\ the grains' optical constants and the grain size distribution,
as well as the dust temperature at the inner boundary; (iii) the relative
thickness of the envelope, i.e. the ratio of outer to inner shell radius,
and the  density distribution; and (iv) the total optical depth at a
given reference wavelength. A two-dimensional approach of the radiative 
transfer problem is applied in, e.g., Men'shchikov et al.\ (2001), 
a hydrodynamic approach with explicit consideration of dust nucleation is
given in Winters et al. (2000). 

Radiative-transfer models have often to rely only on the comparison with the 
observed spectral energy distribution. However, high-resolution spatial 
information 
has proven to be an essential and complementary ingredient of dust-shell
modelling. Only if such information
is available, reliable, i.e.\ non-ambiguous, radiative-transfer models can
be constructed and sound conclusions on the mass-loss process drawn 
(see e.g.\ Bl\"ocker et al.\ 1999, 2001). In the following sections,
we give three examples (IRC+10216, CIT\,3, and IRC+10420) for such modelling.
%
\section{The carbon star IRC+10216}
%
%
The carbon star IRC\,+10\,216 is a long-period  
AGB star suffering from a
strong stellar wind (several $10^{-5}$ M$_{\odot}$/yr; Loup et al. 1993)
which have led to an  almost complete obscuration of the star by dust. 
Due to the high mass-loss rate, long period of $P=649$\,d 
(Le Bertre 1992), and carbon-rich  chemistry of the dust-shell, 
IRC\,+10\,216 is obviously in a very advanced stage of its 
AGB evolution.
High-resolution near-infrared imaging of IRC\,+10\,216
has revealed that on sub-arcsecond scales (100\,mas) 
its dust shell is clumpy, bipolar, 
and changing on a time scale of only $\sim$1\,yr 
(Weigelt et al.\ 1997, 1998,  Haniff \& Buscher 1998,  Osterbart et al.\ 2000, 
 Tuthill et al.\ 2000, Weigelt et al.\ 2002). 
Since most dust shells around AGB stars are known to be spherically 
symmetric, whereas most proto-planetary nebulae (PPN) show an 
axisymmetric geometry (Olofsson 1996), it appears likely that IRC\,+10\,216
has already entered the transition phase to the PPN stage. 
This suggests that the break of the dust-shell symmetry 
between the AGB and post-AGB phase
already takes place at the end of the AGB evolution.

Bispectrum speckle-interferometry observations of IRC\,+10\,216 
were carried out with the SAO 6\,m telescope in the $J$, $H$, and $K$
band  by Osterbart et al.\ (2000) and Weigelt et al.\ (2002) covering eight 
epochs between 1995 and 2001.
Fig.~\ref{FJHKcont} illustrates the different appearance of the dusty 
environment of IRC\,+10\,216 in the $J$, $H$, and $K$ bands.
Fig.~\ref{FKima} shows the reconstructed $K$-band images of the innermost
region of  IRC\,+10\,216 in 1996, 1998 and 2000.
The dust shell consists  
of several compact components, at the beginning within a radius of 200\,mas, 
which steadily change in shape and brightness.
For instance, 
the apparent separation of the two
initially brightest components A and B increased 
from 201 mas in 1996 to 320 mas in 2000.  
At the same time, component B is fading and has almost disappeared in 2000
whereas the initially faint components C and D have become brighter.
In 2001, the intensity level of  component C has increased to almost 40\%
of the peak intensity of  component A.
Both components appear to have started merging in 2000. 

These changes of the dust-shell appearance can be related to 
changes of the optical depths caused, e.g., by mass-loss variations.
The present monitoring, covering more than  3 pulsational 
periods, shows that the structural variations are not related to the 
stellar pulsational cycle in a simple way.
This is consistent with the predictions of 
hydrodynamical models that enhanced dust formation takes place on a timescale
of several pulsational cycles (Fleischer et al.\ 1995).

Recent two-dimensional radiative transfer modelling 
(Men'shchikov et al.\ 2001) has shown 
that the star is  surrounded by an optically
thick dust shell with polar cavities of a full opening angle of 
$36^{\rm o}$, which are inclined by $40^{\rm o}$ pointing with the 
southern lobe towards the observer. The bright and compact
component A is not the direct light from the underlying central star 
but the southern lobe of this bipolar structure dominated
by scattered light. Instead, the carbon star is at the position of the 
fainter northern component B. 
%
%
\section{The oxygen-rich AGB star CIT\,3}
%
%
CIT\,3 is an oxygen-rich
long-period variable star evolving along the AGB with
extreme infrared properties.
Due to substantial mass loss it is surrounded by an optically thick dust shell
which absorbs almost all visible light radiated by the star and
finally re-emits it in the infrared regime.
The first near infrared bispectrum speckle-interferometry
observations of  CIT\,3 in the $J$-, $H$-, and $K^{\prime}$-band
(resolution: 48\,mas, 56\,mas, and 73\,mas)
were obtained with the SAO 6\,m telescope by Hofmann et al.\ (2001).
While CIT\,3 appears almost spherically symmetric in the
$H$- and $K^{\prime}$-band
it is clearly elongated in the $J$-band along a symmetry axis of position angle
$-28^{\rm o}$. Two structures  can be identified: a compact
elliptical core 
and a fainter north-western fan-like structure.
\begin{figure}
\begin{center}
\epsfxsize=10cm
\epsfbox[64 12 550 424]{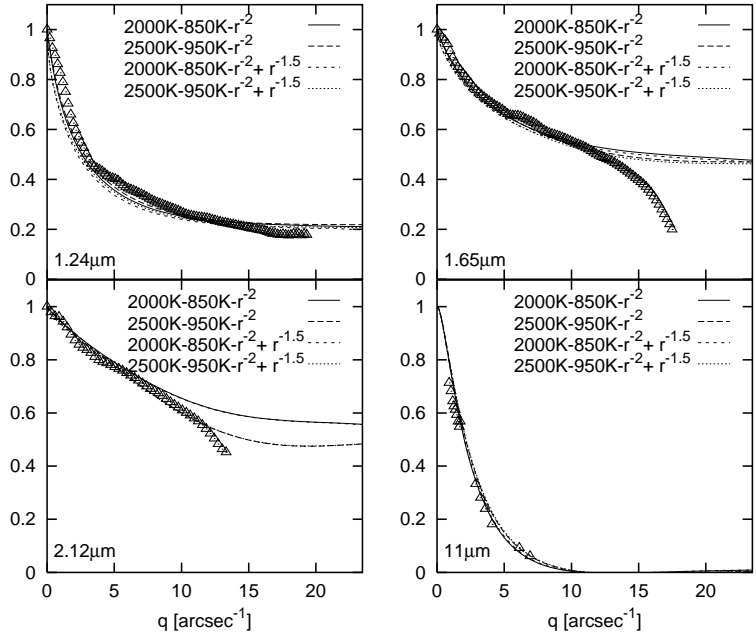}
\end{center}
\caption{CIT\,3 model visibilities at 1.24\,$\mu$m, 1.65\,$\mu$m, 
2.12\,$\mu$m, and 11\,$\mu$m for models with
($T_{\rm eff},T_{1}$)=(2000\,K,850\,K) and (2500\,K,950\,K), resp. 
(Hofmann et al.\ 2001).
Both uniform outflow models ($\rho \sim 1/r^{2}$; solid and long-dahed lines) 
and models with a flatter density distribution in
 the outer shell region
($\rho \sim 1/r^{2}$ for $Y \le  20.5$ and  $\rho \sim 1/r^{1.5}$ for 
$Y > 20.5$; short-dashed and dotted lines) are shown. 
The optical depth $\tau_{0.55\mu{\rm m}}$ is 30. 
The symbols refer to the observations.
Near-infrared visibilities --- SAO 6m speckle observations.
Mid-infrared visibility ---  ISI observations at  4m, 9.6m and 16m baseline
of Lipman et al.\ (2000). For more details of the radiative transfer models
and the fit of the spectral energy distribution, see Hofmann et al.\ (2001).}
\label{FvisiCIT3}
\end{figure}

%
Extensive radiative transfer calculations have been carried out 
and confronted with 
the spectral energy distribution ranging from 1\,$\mu$m to 1\,mm, with the
1.24\,$\mu$m, 1.65\,$\mu$m and 2.12\,$\mu$m  visibility functions, 
as well as with 11$\mu$m ISI interferometry.
The best model found to match the observations refers to a cool central star
with $T_{\rm eff}=2250$\,K which is surrounded by an optically thick 
dust shell with $\tau (0.55\mu m) = 30$ (see Fig.~\ref{FvisiCIT3}).
The central-star diameter is 10.9\,mas
and the inner dust shell diameter 71.9\,mas.
The inner dust-shell rim at $r_{1}= 6.6 R_{\ast}$ has a 
temperature of $T_{1}=900$\,K. 
A two-component model consisting of an inner 
uniform-outflow shell region ($\rho \sim 1/r^{2}$, $r < 20.5 r_{1}$)
and an outer region 
where the density  declines more shallow as $\rho \sim 1/r^{1.5}$
proved to give the best overall match of the observations. 
Provided the outflow velocity stayed constant,
the more shallow density distribution in the outer shell
indicates  that mass-loss has decreased with time in the past of CIT\,3. 
Adopting $v_{\rm exp}=20$\,km/s, the termination of that mass-loss
decrease and 
the begin of the uniform-outflow phase 
took place 87\,yr ago. The present-day mass-loss rate can be determined to be 
$\dot{M} = (1.3-2.1) \cdot 10^{-5}$\,M$_{\odot}$/yr for $d=500-800$\,pc. 
A full description of these observations and models is given in 
Hofmann et al.\ (2001).

CIT\,3 proved to be among the most interesting far-evolved 
AGB stars due to its infrared properties. Moreover, 
the aspherical appearance of its dust shell in the $J$-band 
puts it in one line with the few AGB stars known to expose near-infrared 
asphericities in their dust shells.
The development of such asphericities close to the central
star suggests that CIT\,3 is in the very end of the AGB evolution
or even in transition to the proto-planetary nebula phase where 
most objects are observed in axi\-symmetric geometry (Olofsson 1996).
However, in contrast to other objects (as IRC+10216), CIT\,3 shows these
deviations from spherical symmetry only in the $J$-band,
which is almost completely dominated by scattered light. This suggests that
CIT\,3 had just started to form aspherical structures and is in this regard 
still in the beginning of its final AGB phase. If so, CIT\,3 is one 
of the earliest representatives of this dust-shell transformation phase
known so far.  
\section{The rapidly evolving hypergiant IRC+10420}
The star IRC\,+10\,420
is an outstanding object for the study of stellar evolution
since it is the only object 
currently being observed in its rapid transition from the red supergiant stage
to the Wolf-Rayet phase.
Its spectral type changed from
F8\,I$_{\rm a}^{+}$ in 1973 (Humphreys et al.\ 1973) 
to mid-A today (Oudmaijer et al.\ 1996)
corresponding to an increase of its effective temperature 
of 1000-2000\,K within only 25\,yr.
It is heavily obscured by circumstellar dust due to strong mass loss with
rates typically of the order of several $10^{-4}$\,M$_{\odot}$/yr.
IRC\,+10\,420 can be classified as a luminous hypergiant with a
mass  of initially $\sim 20$ to 40 M$_{\odot}$.

Diffraction-limited  73\,mas bispectrum speckle interferometry of 
IRC\,+10\,420 (Bl\"ocker et al.\ 1999) shows that 
the $K$-band visibility drops to 0.6 and then stays constant 
for frequencies $>4$\,cycles/arcsec revealing that 
the central star contributes $\sim$ 60\% and the dust shell $\sim$
40\% to the total flux.  To interpret these observations in more detail,
radiative transfer calculations were conducted 
taking into account both SED and visibility. 
Again, single-shell models failed to reproduce the observations
and  a two-component shell was introduced
assuming that IRC\,+10\,420 had passed through a
superwind phase in its history as can be expected from its evolutionary
status.  A previous superwind phase leads to changes in the density
distribution, i.e.\ there is a region in the dusty shell which shows a
density enhancement over the normal $r^{-2}$ distribution.  The best
model for both SED and visibility was found for a dust shell
with a dust temperature of 1000 K at its inner radius of $r_{1}=69 R_{\ast}$.
At a distance of $308 R_{\ast}$ ($Y=r/r_{1}=4.5$), 
where the dust temperature has dropped
to 480 K, 
the density was enhanced by a factor of $S=40$ and its slope within the shell
changed from $1/r^{2}$ to $1/r^{1.7}$.
The angular
diameters of these components are 69 mas and 311 mas (stellar diameter
$\sim$ 1 mas for $d=5$ kpc).  This can be interpreted in terms of a 
termination of an 
enhanced mass-loss phase roughly 90 years ago.  The mass-loss rates of
the components can be determined to be
$\dot{M}_{1}= 7.0\ 10^{-5}$\,$M_{\odot}/{\rm yr}$ and
$\dot{M}_{2}= 1.1\ 10^{-3}$\,$M_{\odot}/{\rm yr}$.
We refer to Bl\"ocker et al.\ (\cite{BloeEtal99})
for a full description of the model grid. 

%
\section{Simulating VLTI observations of IRC+10420}
\begin{figure}
\begin{center}
\epsfxsize=0.80\textwidth
\epsfbox[39 265 403 661]{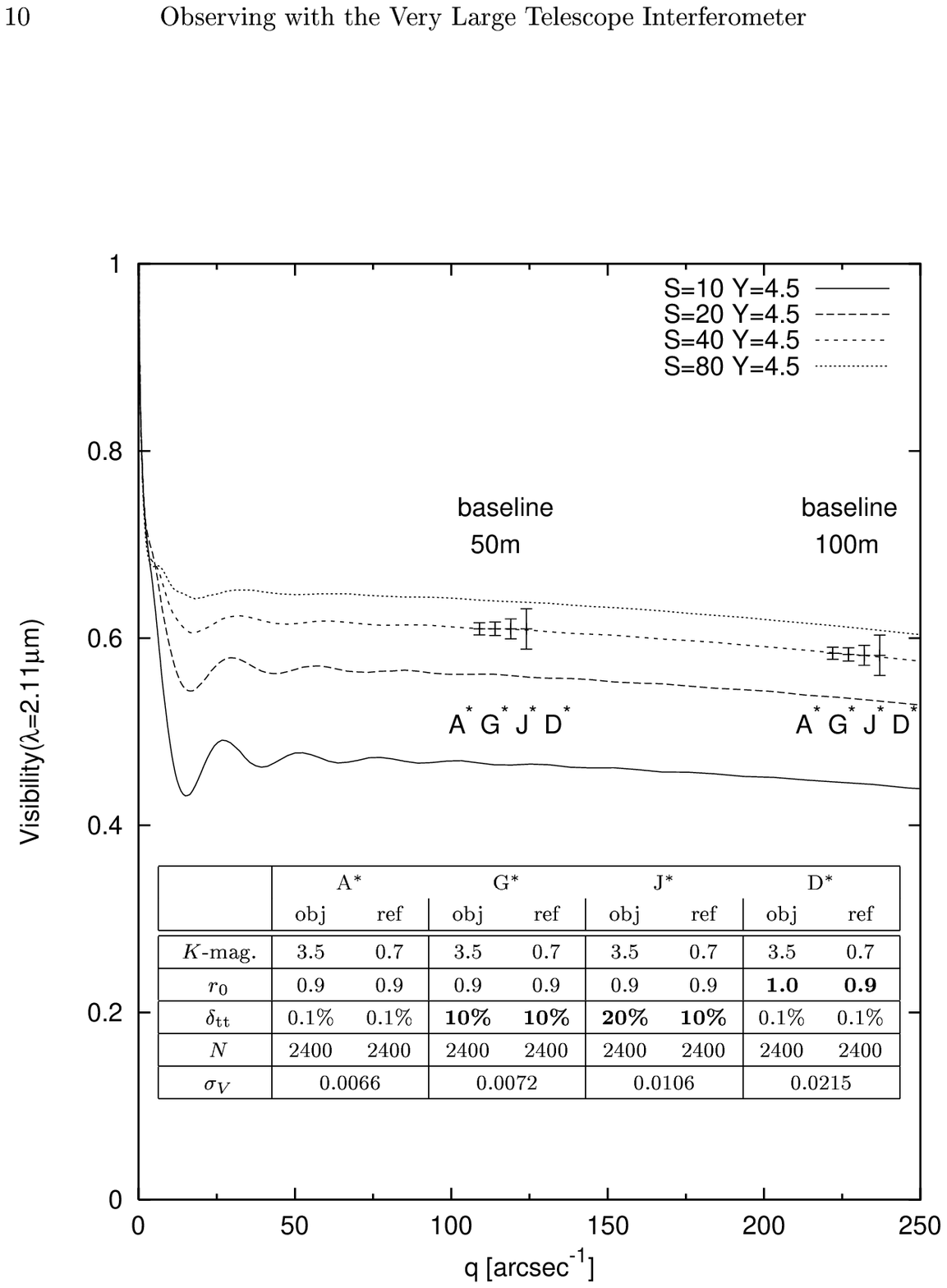}
\end{center}
\vspace*{-2mm}
\caption{Dependence of the error bars of simulated AT-VLTI/AMBER 
(wide-field mode) observations of IRC\,+10\,420 at 2.11\,$\mu$m 
(Przygodda et al. 2001)
on (i) seeing  differences of object and reference star
observations (Fried parameter, $r_{0}$, differences; 
simulation D$^{\ast}$), and
(ii) different residual tip-tilt errors, $\delta_{\rm tt}$,
of object and reference star observations 
(simulations G$^{\ast}$ and J$^{\ast}$). 
Lines refer to radiative transfer models of
different superwind amplitudes $S$. The symbols refer to 
AT-VLTI/AMBER simulations (each with $N$=2400 interferograms)
of the $S$=40 intensity profile
for baselines of 50 and 100\,m (100\,m: $q$=227 arc\,sec$^{-1}$).
To better distinguish between the simulations,
the data points belonging to one baseline are somewhat shifted 
with respect to the spatial frequency. 
The error bars are based on  6 statistically independent repetitions of
each simulation. 
} \label{FsimvisATH}
\end{figure}

%
%
The above data (and model) of IRC+10420
rely on observations with the SAO 6\,m telescope.
Fig.~\ref{FsimvisATH} shows the visibility model predictions for different 
superwind amplitudes $S$ up to a baseline of 110\,m. Obviously, the various
models can be best distinguished at longer baselines. 
Taking the corresponding model intensity distributions as input, 
Prygodda et al.\ (2001) presented 
computer simulations of interferometric imaging with
VLTI/AMBER (ATs, wide-field mode) for IRC+10420.
%
These simulations consider light propagation from the object to the
detector as well as photon noise and detector read-out noise
and show the dependence of the
visibility error bar on various observational parameters. The results are 
shown in Fig.~\ref{FsimvisATH}.
Different seeing conditions 
for object and reference star turn out to be
more crucial than, e.g., residual tip-tilt errors.
With these simulations at hand one can immediately see under
which conditions the visibility data quality would allow us to discriminate
between the different model assumptions (here: the size of the superwind 
amplitude $S$).
Inspection of
Fig.~\ref{FsimvisATH} shows that in all studied cases 
the observations will give clear preference to one particular model.
Therefore, observations with VLTI will certainly be well suited to 
prove theroretical predictions and to improve our current knowledge 
of this outstanding object.
%
%
\section{Evolved Stars and the VLTI}
%
%
VLTI observations with AMBER (see Petrov, this volume) and 
MIDI (see Perrin, this volume) 
will certainly have a large impact on the study of 
stars in late stages of stellar evolution revealing, for example, details
of dust-shell structures and the mass-loss process. 
Radiative transfer 
calculations are an appropriate and efficient tool of interpreting these 
observations. 
However, for the robust and non-ambiguous construction of dust-shell
models it is essential to take diverse and independent observational
constraints into account. 
Apart from matching the spectral energy distribution, the 
consideration of spatially resolved information plays a crucial role for
obtaining a reliable model. Generally speaking, as many pieces of 
spectral information as possible have to be taken into account. 
Visibilities at various wavelengths greatly constrain modelling, probing, 
for instance, scattering ($J$ band) and  
thermal emission of hot dust ($H$, $K$ band) and cool dust ($N$ band).
Near infrared visibilities can serve as sensitive indicators of the grain size.
Furthermore, the combination of long-baseline interferometry data with 
high-resolution data at short baselines, as obtained, e.g., by 
bispectrum speckle interfermetry,
will provide complementary information and will be of utmost value for  
modelling and interpretation (see, e.g., the case of IRC+10420).

For example, a near-infrared survey of
dusty AGB stars with  VLTI-AMBER should be based on 
a combination of various selection criteria
comprising 
the existence of high-resolution single-dish observations, 
and/or prominent location in
color-color (e.g.\ $J{-}K$ vs.\ $K$-$[12]$) and magnitude-color  
(e.g.\ $K$ vs. $K$-$[12]$) diagrams indicative for the presence of dust, 
i.e. close vicinity to already resolved dusty objects. 
Dusty AGB objects resolved by speckle interfero\-metry mostly 
show  $K{-}[12]\ga$5. Due to their thick dust shells, most of these stars are 
bright in $K$ and observable with the VLTI without fringe tracking. 
%
%
%
%
%
%
%


\begin{thebibliography}{99}
%
%
%

\bibitem[2001]{BloeEtal01}
 Bl\"ocker T., Balega Y., Hofmann K.-H., Weigelt G., 2001, A\&A 369, 142

\bibitem[1999]{BloeEtal99}
 Bl\"ocker T., Balega Y., Hofmann K.-H., Lichtenth\"aler J., Osterbart R.,
 Weigelt G., 1999, A\&A 348, 805

\bibitem[2002]{BranEtal02}
 Brandner W., Rousset G., Leuzen R., Hubin N., Lacombe F., Hofmann R., et al.,
 2002, Msngr 107, 1

\bibitem[1995]{FleiEtal95}
 Fleischer A.J., Gauger A., Sedlmayr E.,  1995, A\&A 297, 543

\bibitem[1998]{HanBus98}
 Haniff C.A., Buscher D.F., 1998, A\&A 334, L5

\bibitem[1986]{HofWei86}
 Hofmann K.-H., Weigelt G., 1986, A\&A 167, L15

%
%
\bibitem[2001]{HofEtal01}
 Hofmann K.-H., Bl\"ocker T., Balega Y., Weigelt G., 2001, A\&A 379, 529

\bibitem[1973]{HumStrMurLow73} 
 Humphreys R.M., Strecker D.W., Murdock T.L., Low, F.J., 1973, ApJ 179, L49

\bibitem[1997]{IveEli97}
    Ivezi\'c \v{Z}., Elitzur M., 1997, MNRAS 287, 799

%
\bibitem[1992]{LeB92}
 Le Bertre T., 1992, A\&AS 94, 377

\bibitem[2000]{LipEtal2000}
 Lipman E.A., Hale D.D., Monnier J.D., Tuthill P.G., Danchi W.C., Townes C.H.,
 2000, ApJ 532, 467

\bibitem[1983]{LohWeiWir83}
 Lohmann A.W., Weigelt G., Wirnitzer B., 1983, Appl. Opt. 22, 4028

\bibitem[1993]{LoupEtal93}
 Loup C., Forveille T., Omont A., Paul J.F., 1993, A\&AS 99, 291. 

\bibitem[2001]{MenEtal01}
 Men'shchikov A., Balega Y., Bl{\"o}cker T., Osterbart R., \& Weigelt
  G. 2001, A\&A 368, 497 

%
\bibitem[1996]{Olof96}
 Olofsson H., 1996, ApSS 245, 169 

\bibitem[2000]{OstEtal00}
 Osterbart R., Balega Y., Bl{\"o}cker T., Men'shchikov A., Weigelt G.,
 2000, A\&A 357, 169 

\bibitem[1996]{OudGroeMatBloSah96}
    Oudmaijer R.D., Groenewegen M.A.T., Matthews H.E., Blommaert J.A.D.L,
    Sahu K.C., 1996, MNRAS 280, 1062

\bibitem[2001]{PrzEtal01}
 Przygodda F.,  Bl{\"o}cker T., Hofmann K.-H., Weigelt G., 2001, Opt.Eng. 40,
 753

\bibitem[2000]{TutEtal00}
 Tuthill P.G., Monnier J.D., Danchi W.C., Lopez B., 2000, ApJ 543, 284

\bibitem[2002]{TysEtal02}
 Tyson R.K., Bonaccini D., Roggemann M.C. (eds.), 2002, Adaptive Optics Systems
  and Technology II, Proc. SPIE Vol.\ 4494

\bibitem[1977]{Wei77}
 Weigelt G., 1977, Optics Commun. 21, 55

%
\bibitem[1997]{WeiEtal97}
  Weigelt G., Balega, Y., Hofmann, K.-H., Langer, N., Osterbart, R., 1997,
  Science with the VLT Interferometer, ESO Astrophysics Symposia, p. 206  

\bibitem[1998]{WeiEtal98}
  Weigelt G., Balega Y., Bl{\"o}cker T., Fleischer A.J., Osterbart R.,
  Winters J.M., 1998, A\&A 333, L51 

\bibitem[2002]{WeiEal02}
  Weigelt G., Balega Y., Bl{\"o}cker T., Hofmann K.-H., Men'shchikov A., 
  Winters J.M., 2002, {\it Planetary Nebulae}, IAU Symp.\ 209, 
 M.\ Dopita, S.\ Kwok, and R.S. Sutherland (eds.), Astronomical Society of the 
 Pacific, in press 

\bibitem[2000]{WinEtal00}
  Winters J.M., Le Bertre T., Jeong K.S., Helling C., Selmayr E., 2000, 
  A\&A 361, 641 
%
\end{thebibliography}
\end{document}